\documentclass[final,5p,times,twocolumn]{elsarticle}

\usepackage{enumerate}
\usepackage{amsmath}
\usepackage{amsfonts}
\usepackage{amssymb}
\usepackage[utf8]{inputenc}
\usepackage[T1]{fontenc}
\usepackage{mathtools}
\usepackage{wasysym}
\usepackage{accents}
\usepackage[svgnames]{xcolor}
\usepackage[colorlinks=true, pdfstartview=FitV, linkcolor=blue, citecolor=red, urlcolor=magenta]{hyperref}
\usepackage{url}
\usepackage{graphicx}
\usepackage{subfig}
\usepackage{bm}
\usepackage{tikz}
\usetikzlibrary{arrows,decorations.markings}
\usepackage{amsmath}
\usepackage{widetext}

\newtheorem{theorem}{Theorem}[section]

\newtheorem{definition}{Definition}[section]

\journal{Physics Letters B}

\begin{document}
\begin{frontmatter}

\newcommand{\Areia}{
\address{Department of Chemistry and Physics, Federal University of Para\'iba, \\ Rodovia BR 079 - Km 12, 58397-000 Areia-PB,  Brazil.}
}
\newcommand{\Lavras}{
\address{Physics Department, Federal University of Lavras, Caixa Postal 3037, 37200-000 Lavras-MG, Brazil.}
}

\newcommand{\UFRJ}{
\address{Instituto de F\'isica, Universidade Federal do Rio de Janeiro, 21.941-972 - Rio de Janeiro-RJ - Brazil}
}

\newcommand{\PII}{
\address{Departamento de F\'isica and Mestrado Profissional em Pr\'aticas de Educa\c c\~ao B\'asica (MPPEB),\\ Col\'egio Pedro II, 20.921-903 - Rio de Janeiro-RJ - Brazil}
}

\title{A varying gravitational constant map in asymptotically AdS black hole thermodynamics}

\author{Iarley P. Lobo}\ead{lobofisica@gmail.com}
\Areia
\Lavras

\author{Jo\~ao Paulo Morais Gra\c ca}\ead{jpmorais@gmail.com}
\UFRJ

\author{Eduardo Folco Capossoli}\ead{eduardo\_capossoli@cp2.g12.br}
\PII

\author{Henrique Boschi-Filho}\ead{boschi@if.ufrj.br}
\UFRJ

\date{\today}

\begin{abstract}
We propose a sequence of steps and a generic transformation for connecting common thermodynamic quantities considered in asymptotically anti-de Sitter black hole thermodynamics in the bulk and those that are appropriate for CFT thermodynamics in the boundary. We do this by constructing a ``varying-$G$ map'', where $G$ is the gravitational constant,  and demonstrate its usefulness by considering various examples.
\end{abstract}

\end{frontmatter}


\section{Introduction}
 In 1997, Maldacena proposed a duality between string theories in an anti-de Sitter (AdS) spacetime (the bulk) and conformal field theories at its boundary \cite{Maldacena:1997re}, which could give insights, for instance, about a theory of quantum gravity \cite{Addazi:2021xuf}. Based on this idea, Witten related phase transitions in the bulk with the confinement/deconfinement transition of the gauge fields at the boundary \cite{Witten:1998zw}. The phase transitions considered by Witten are the so-called Hawking-Page phase transitions, originally the transition between a black hole in an anti-de Sitter space and pure radiation in this same space \cite{Hawking:1982dh}. In thermodynamics, we say this is a first-order transition.

Witten's analysis is based on traditional black hole thermodynamics, where the black hole's mass, for example, corresponds to its internal energy. More recently, however, Kastor, Ray and Traschen have proposed a modification to this framework, equating the black hole's mass not with its internal energy, but with its enthalpy \cite{Kastor:2009wy}. With that, they were able to introduce a notion of a mechanical pressure for the black hole, as well as a thermodynamic volume.

In this new framework, Kubiznak and Mann showed that anti-de Sitter charged black holes exhibit other types of phase transitions, such as second-order phase transitions, similar to the phase transitions of usual matter, which often can be described by equations of state such as that of Van der Waals \cite{Kubiznak:2012wp}. Because the black hole's mass is equaled to its enthalpy, such a framework was called black hole chemistry, to differentiate it from the usual thermodynamics of black holes (for a review on the subject, see \cite{Kubiznak:2016qmn}). This new framework, however, requires extensions to the dictionary proposed by Witten. The reason for this is multiple. For instance, the mechanical pressure of the black hole (the pressure in the bulk) is not the same as the boundary pressure \cite{Johnson:2014yja}. But what does this pressure mean then? And how to calculate the boundary pressure from the bulk pressure?

Another key point is the relationship between the size of the anti-de Sitter space and the number of colors, $N$, of the dual Yang-Mills (Y-M) theory or degrees of freedom of the Conformal Field Theory (CFT) at the boundary. According to the usual dictionary, the latter is measured by the central charge, $C$, as $N^2 \propto C\propto L^{D-2}/G$, where $L$ is the anti-de Sitter radius, $G$ is the gravitational constant in $D$ bulk dimensions \cite{Visser:2021eqk,Cong:2021fnf}. This means that, for a fixed $G$, a change in the size of the anti-de Sitter space leads to a variation in the number of colors, i.e., we modify the Y-M theory at the boundary or takes us from a CFT to another.
\par
On the other hand, the radius of the AdS space is related to the volume of the CFT as ${\cal V}=L^{D-2}$. Therefore if one aims to vary, in an independent way, both the volume and the central charge (or degrees of freedom) of the CFT, it becomes necessary to allow the gravitational constant to vary. Or if one aims to consider modifications of the CFT volume, but keeping the same theory, it becomes necessary to also vary simultaneously the gravitational constant and the AdS radius in order to keep the central charge unmodified (for further details, please refer to \cite{Visser:2021eqk} and references therein). It should be emphasized that had one not considered a variation of the AdS radius (or equivalently the cosmological constant), Witten's dictionary would be perfectly fine, and the problematic behavior mentioned above emerges in the context of the extended black hole thermodynamics, that introduces a mechanical pressure due to a variation of the cosmological constant.
\par
The compatibility between the black hole thermodynamics valid in the bulk and the CFT at the boundary requires that a dictionary is constructed in a way that relates the ``hair'' parameters that define the black hole and the CFT thermodynamic quantities. Usually, this is done by identifying a map between the boundary-defined quantities and the bulk black hole ones in a way that preserves the first law and Smarr relations in both representations. Recently, such an approach has been carried out for cases with rotation \cite{Visser:2021eqk,Cong:2021fnf}, electric charge with different types of electrodynamics (reflected in the black hole solutions considered) \cite{Cong:2021jgb,Kumar:2022fyq,Rafiee:2021hyj}, and also for string theory corrections modeled by higher derivative terms \cite{Dutta:2022wbh} in a way that present, as a common point, the need for the aforementioned redefinition of variables for each case under scrutiny.

In this paper, we construct a general algorithm for identifying the necessary transformation between generic thermodynamic quantities that are considered in standard black hole thermodynamics in the bulk and those needed for making this thermodynamics compatible with the one carried out in the boundary CFT. The dictionary itself is written in definition \ref{def:dict} and the proof of theorem \ref{theo:1law} (that states the compatibility between the black hole thermodynamics in the bulk and the one in the boundary) is done throughout section \ref{sec:dict}.  Using this algorithm (that is explicitly written in section \ref{sec:examples}), we show its application by recovering various known results in the literature and we propose how it should be applied in a novel scenario. Finally, we draw our final remarks in section \ref{sec:conc}.


\section{Varying-$G$ map}\label{sec:dict}
The starting point of this derivation consists in identifying how the gravitational constant $G$ scales in $D$ spacetime dimensions. In fact, for simplicity, assuming units in which $c=\hbar=1$ (which can be assumed since these quantities are held fixed), if $l$ is a fiducial quantity with length dimensions, we have that the dimension of $G$ behaves like $[G]=l^{D-2}$ \cite{Cong:2021fnf}. Given a set of thermodynamic quantities $\tau_i$ $(i=1,...,n)$ which are not coupled to the gravitational constant $G$, let us define a set of new thermodynamic variables $\eta_i=G^{\alpha_i}\tau_i$ (no sum over indices assumed here).
The motivation for this definition stems from usual couplings of $\alpha_i$ powers of $G$ and $\tau_i$ in the metric functions of black holes, which will serve as the starting point of the holographic map that we shall propose.\footnote{
To see how this works, consider  the Reissner-Nordstr\"{o}m metric function \cite{Alfaia:2021cnk} $f(r)=1-\frac{2 GM}{r^{D-3}}+\frac{GQ^2}{r^{2(D-3)}}$, where $M$ and $Q$ are the black hole's mass and charge. In this case, by choosing $\eta_1=GM$ and $\tau_1=M$, we have $\alpha_1=1$. Besides that, if $\eta_2=\sqrt{G}Q$ and $\tau_2=Q$, we have $\alpha_2=1/2$. This way, the metric becomes
$f(r)=1-\frac{2\eta_1}{r^{D-3}}+\frac{\eta_2^2}{r^{2(D-3)}}$.}

The length dimension of the variables $\eta_i$ scale as $[\eta_i]=l^{\alpha_i(D-2)}[\tau_i]$ 
and if $\tau_i$ scale as $[\tau_i]=l^{\beta_i}$, then we have 
\begin{equation}
[\eta_i]=l^{\alpha_i(D-2)+\beta_i}\, .
\end{equation}
The thermodynamic quantities $\eta_i$ have conjugate potentials $A_i$, which can also present factors of $G$ once one redefines the $\tau_i$ variables, but we shall not deal with them, because the roles of $\tau_i$ and $A_i$ only differ by a Legendre transformation.

At this point, one should notice that although, for simplicity, we are using an auxiliary quantity $l$ to describe length dimensions, what we have, in fact, is an implicit scaling of different thermodynamic quantities with $G$, since $l=[G]^{1/(D-2)}$.
\par
Now, suppose that we consider a black hole whose mechanic/thermodynamic variables are the redefined mass $GM$, a term proportional to the area as $A/4$ (with conjugate variable proportional to the surface gravity as $\kappa/2\pi$), a negative cosmological constant $\Lambda/8\pi$ (with a potential $\Theta$) and a set of variables $G^{\alpha_i}\tau_i$ (and related potentials $A_i$).
\par
Using the fact that these quantities scale as $[G]=l^{D-2}$, $[M]=l^{-1}$, $[A]=l^{D-2}$, $[\Lambda]=l^{-2}$ and $[G^{\alpha_i}\tau_i]=l^{(D-2)\alpha_i+\beta_i}$, we have the following first law of thermodynamics and Smarr relation, that follows from Euler's theorem for homogeneous functions (where we have $\sum_i\doteq\sum_{i=1}^n$):
\begin{align}
d(GM)&=\frac{\kappa}{8\pi}dA+\frac{\Theta}{8\pi}d\Lambda+\sum_iA_id(G^{\alpha_i}\tau_i)\, ,\label{1law-eta}\\
    (D-3)M&=(D-2)\frac{\kappa A}{8\pi G}-\frac{\Theta \Lambda}{4\pi G}\nonumber \\
    &\quad+\sum_i[(D-2)\alpha_i+\beta_i]G^{\alpha_i-1}\tau_iA_i\, .\label{smarr0}
\end{align}

These expressions are just the usual first law and Smarr relation of black hole thermodynamics and for quantities like entropy $S=A/4G$, temperature $T=\kappa/2\pi$, black hole's chemistry pressure $P=-\Lambda/8\pi G$ and volume $V=-\Theta$. For example, by setting $n=1$ and $\tau_1=Q$, $A_1=\sqrt{G}\Phi$, $\alpha_1=1/2$ and $\beta_i=(D-4)/2$, we are able to derive the usual Smarr relation of the Reissner-Nordstr\"{o}m-anti-de Sitter (AdS) black hole in $D$-dimensions $(D-3)M=(D-2)TS+2PV+(D-3)Q\Phi$ (please, refer to Ref.\cite{Cong:2021fnf,Alfaia:2021cnk} to verify the scaling of these quantities.)

As described in \cite{Visser:2021eqk} and in the introduction of our paper, the compatibility of these expressions with the holographic principle leads to a necessary variation of the gravitational constant. In this case, the above first law leads to 
\begin{align}
    dM&=\frac{\kappa}{8\pi}d\left(\frac{A}{G}\right)+\frac{\kappa A}{8\pi G}\frac{dG}{G}+\frac{\Theta}{8\pi G}d\Lambda\nonumber\\ &\quad+ \sum_i\left[A_id\left(\frac{G^{\alpha_i}\tau_i}{G}\right)+\frac{A_i}{G}G^{\alpha_i}\tau_i\frac{dG}{G}\right]-M\frac{dG}{G}\, ,\quad 
\end{align}
where we integrated by parts the terms involving the area $A$ and the parameters $\tau_i$. Rearranging this expression, we can isolate the term that depends on the variation of $G$
\begin{align}\label{1law-dG}
    dM&=\frac{\kappa}{8\pi}d\left(\frac{A}{G}\right)+\frac{\Theta}{8\pi G}d\Lambda + \sum_iA_i d\left(G^{\alpha_i-1}\tau_i\right)\nonumber\\
    &\quad+\frac{dG}{G}\left(-M+\frac{\kappa A}{8\pi G}+\sum_iA_iG^{\alpha_i-1}\tau_i\right)\, .
\end{align}
The choice of the remaining term with $G^{\alpha_i-1}\tau_1$ will be important for a future simplification. 
\par
We now redefine the negative cosmological constant in terms of the AdS radius $L$, as follows \cite{Cong:2021fnf}
\begin{equation}\label{lambda1}
    \Lambda=-\frac{(D-1)(D-2)}{2L^2}\, .
\end{equation}

Using the Smarr relation \eqref{smarr0}, we express $\Theta/(8\pi G)$ as a function of the other variables, 
which can be used in Eq.\eqref{1law-dG}, along with the relation $d\Lambda/\Lambda=-2\, dL/L$ (which follows from \eqref{lambda1}) to furnish
\begin{widetext}
\begin{align}
    dM&=\frac{\kappa}{8\pi}d\left(\frac{A}{G}\right)+ \sum_iA_id(G^{\alpha_i-1}\tau_i)+\frac{dG}{G}\left(-M+\frac{\kappa A}{8\pi G}+\sum_iA_iG^{\alpha_i-1}\tau_i\right)\nonumber\\
  &\quad-\frac{dL}{L}\left[-(D-3)M +(D-2)\frac{\kappa A}{8\pi G}+\sum_i\left[(D-2)\alpha_i+\beta_i\right]G^{\alpha_i-1}\tau_iA_i\right].
\end{align}

Now, we express the variation of $G$ in term of the variation of novel variables $L^{D-2}$ and $L^{D-2}/G$ (which will be related to the CFT volume and central charge). This gives
\begin{align}
    \frac{dG}{G}=\frac{dL^{D-2}}{L^{D-2}}-\frac{d(L^{D-2}/G)}{L^{D-2}/G}\, , \qquad     \frac{dL}{L}=\frac{1}{D-2}\frac{dL^{D-2}}{L^{D-2}}\, .
\end{align}

This leads us to derive the following expressions for the 1st law, that we express in details below:
\begin{align}
    dM&=\frac{\kappa}{8\pi}d\left(\frac{A}{G}\right)+\left(\frac{dL^{D-2}}{L^{D-2}}-\frac{d(L^{D-2}/G)}{L^{D-2}/G}\right)\left(-M+\frac{\kappa A}{8\pi G}+\sum_iA_iG^{\alpha_i-1}\tau_i\right)+\sum_iA_id(G^{\alpha_i-1}\tau_i)\nonumber\\
    &\quad-\frac{1}{D-2}\frac{dL^{D-2}}{L^{D-2}}\left[-(D-3)M+(D-2)\frac{\kappa A}{8\pi G}+\sum_i\left[(D-2)\alpha_i+\beta_i\right]G^{\alpha_i-1}\tau_iA_i\right]\nonumber\\
    &=\frac{\kappa}{8\pi}d\left(\frac{A}{G}\right)+ \sum_iA_id(G^{\alpha_i-1}\tau_i)+\frac{d(L^{D-2}/G)}{L^{D-2}/G}\left(M-\frac{\kappa A}{8\pi G}-\sum_iA_iG^{\alpha_i-1}\tau_i\right)\nonumber\\
  &\quad+  \frac{dL^{D-2}}{L^{D-2}}\left(-\frac{M}{D-2}+\sum_i\left(1-\alpha_i-\frac{\beta_i}{D-2}\right)G^{\alpha_i-1}A_i\tau_i\right)\nonumber\\
  &=\frac{\kappa}{8\pi}d\left(\frac{A}{G}\right)+ \sum_iA_id(G^{\alpha_i-1}\tau_i)+\frac{d(L^{D-2}/G)}{L^{D-2}/G}\left(M-\frac{\kappa A}{8\pi G}-\sum_iA_iG^{\alpha_i-1}\tau_i\right)\label{dict1}\\
   &\quad-\frac{M}{D-2}\frac{dL^{D-2}}{L^{D-2}}+(D-2)\frac{dL}{L}\sum_i\left(1-\alpha_i-\frac{\beta_i}{D-2}\right)G^{\alpha_i-1}\tau_iA_i\, .\nonumber
\end{align}

In the last step, we used again $dL^{D-2}/L^{D-2}=(D-2)dL/L$. We can rearrange again this equation in terms of new variables $\tau_iG^{\alpha_i-1}L^{(D-2)(1-\alpha_i)-\beta_i}$ and $A_iL^{(D-2)(1-\alpha_i)-\beta_i}$ by joining the second and the last terms of the above equation. This leads to
\begin{align}\label{dict2}
    dM&=\frac{\kappa}{8\pi}d\left(\frac{A}{G}\right)+ \sum_i\frac{A_i}{L^{(D-2)(1-\alpha_i)-\beta_i}}d\left(\tau_iG^{\alpha_i-1}L^{(D-2)(1-\alpha_i)-\beta_i}\right)
    -\frac{M}{D-2}\frac{dL^{D-2}}{L^{D-2}}
    \nonumber\\
    &\quad 
    +\frac{d(L^{D-2}/G)}{L^{D-2}/G}\left(M-\frac{\kappa A}{8\pi G}-\sum_i\frac{A_i}{L^{(D-2)(1-\alpha_i)-\beta_i}}\tau_iG^{\alpha_i-1}L^{(D-2)(1-\alpha_i)-\beta_i}\right)\, .
\end{align}
\end{widetext}

At this point, we are ready to derive the ``varying-$G$ map'' for any set of variables. 

\begin{definition}[Varying-$G$ map]\label{def:dict}
Let $M$, $A$, $\kappa$ and $L$ be the black hole's mass, area, surface gravity and anti-de Sitter radius in $D$ spacetime dimensions. Consider a set of $n$ thermodynamic quantities, called charges, $\tau_i$ (where $n=1,...,n$), that scale with length dimensions $l^{\beta_i}$ and that couple to the $\alpha_i$-th power of the gravitational constant (which presents length dimension $l^{D-2}$) as $\tau_iG^{\alpha_i}$, where $l$ is fiducial length scale. The varying-$G$ map consists in the transformation from the set of thermodynamic variables $(M,A,\kappa,L,\tau_i,A_i,G)$ to the set $(E,S,T,{\cal V},p,\nu_i,B_i,C,\mu)$ as follows
\begin{align}
E&=M\, ,\quad S=\frac{A}{4G}\, ,\quad T=\kappa/2\pi\, ,\label{energy1}\\
{\cal V}&=L^{D-2}\, ,\quad p=\frac{E}{{(D-2)\cal V}}\, ,\label{v1}\\
\nu_i&=\tau_iG^{\alpha_i-1}L^{(D-2)(1-\alpha_i)-\beta_i}\, ,\label{nui}\\
B_i&=\frac{A_i}{L^{(D-2)(1-\alpha_i)-\beta_i}}\, \label{bi}\\
C&=\frac{k}{16\pi} \frac{L^{D-2}}{G} \, , \\ 
\mu&= \frac{1}{C}(E-TS-\sum_iB_i\nu_i)\, ,\label{mu1}
\end{align}
where $k$ is an arbitrary constant that labels different holographic systems. 
The set $(E,S,T,{\cal V},p,C,\mu)$ corresponds to variables known as internal energy, entropy, temperature, volume, pressure, central charge and chemical potential, respectively. The quantities $(\nu_i,B_i)$ are the redefined charges and potentials.
\end{definition}

\begin{theorem}\label{theo:1law}
Let $M$, $A$, $\kappa$, $\Lambda=-(D-1)(D-2)/(2L^2)$ and $\Theta$ be the black hole's mass, area, surface gravity, cosmological constant and the conjugate variable to $\Lambda$ in $D$ spacetime dimensions, respectively (where $L$ is the anti-de Sitter radius). Consider a set of $n$ thermodynamic quantities $\tau_i$ (where $n=1,...,n$) that scale with length dimensions $l^{\beta_i}$ and that couple to the $\alpha_i$-th power of the gravitational constant as $\tau_iG^{\alpha_i}$. If $G$ is a variable gravitational constant (that presents length dimension $l^{D-2}$), such that the black hole's first law and Smarr relation read
\begin{align}
  d(GM)&=\frac{\kappa}{8\pi}dA+\frac{\Theta}{8\pi}d\Lambda+\sum_iA_id(G^{\alpha_i}\tau_i)\, ,\label{1stlaw1}\\
  (D-3)M&=(D-2)\frac{\kappa A}{8\pi G}-\frac{\Theta \Lambda}{4\pi G}+\sum_i[(D-2)\alpha_i+\beta_i]G^{\alpha_i-1}\tau_iA_i\, ,\label{smarr1}
\end{align}
then the varying-$G$ map consists in a transformation between these expressions and the following holographic first law and holographic Smarr relation
\begin{align}
    dE&= TdS+\sum_i B_id\nu_i-pd{\cal V}+\mu dC\, ,\label{1stlaw2}\\
    E&= TS+\sum_iB_i\nu_i+\mu C\, .\label{smarr2}
\end{align}
\end{theorem}

The substitution of Eqs.\eqref{energy1}-\eqref{mu1} into equation \eqref{dict2} constitutes a proof of the above theorem.

This way, from the holographic first law, one sees that the quantity $E$ is the equivalent of an internal energy, $S$ and $T$ are the entropy and temperature, $p$ and ${\cal V}$ are the pressure and volume. Since the central charge $C=kL^{D-2}/(16\pi G)$ measures the number of degrees of freedom of a corresponding CFT
in the AdS/CFT correspondence, then the quantity $\mu$ should be interpreted as a chemical potential in this analogy. The rest of the varying-$G$ map, i.e., that defines $\nu_i$ and $B_i$, corresponds to the appropriate choice of variables to extend the holographic principle to black hole thermodynamics.


\section{General algorithm and examples}\label{sec:examples}
A general algorithm for applying the varying-$G$ map would be the following:
\begin{enumerate}
\item Identify the relevant charges, $\tau_i$, in which the standard black hole thermodynamics would be formulated.
\item Explicitly write $G$ in the equations of the gravitational system (for example, in the metric), in order to identify the power, $\alpha_i$, in which it couples to $\tau_i$ for finding the quantity $\eta_i=G^{\alpha_i}\tau_i$.
\item Discover the power length dimensions, $\beta_i$, of $\tau_i$ as $[\tau_i]=l^{\beta_i}$.
\item Express the first law of black hole thermodynamics through variations of $GM$, $A$, $\Lambda$ and $G^{\alpha_i}\tau_i$ as
\begin{eqnarray}\label{1law-eta2}
 d(GM)=\frac{\kappa}{8\pi}dA+\frac{\Theta}{8\pi}d\Lambda+\sum_iA_id(G^{\alpha_i}\tau_i)\, .
\end{eqnarray}
\item From the expected form of the Smarr relation in standard black hole thermodynamics
\begin{equation}\label{smarr-eta2}
    (D-3)M=(D-2)\frac{\kappa A}{8\pi G}-\frac{\Theta \Lambda}{4\pi G}+\sum_i[(D-2)\alpha_i+\beta_i]G^{\alpha_i-1}\tau_iA_i\, .
\end{equation}
one can find the form of the potentials $A_i$ as given by the product of $\gamma_i$-powers of $G$ and the usual potentials $\Phi_i$ that are conjugate to $\tau_i$ in standard black hole thermodynamics (the one in which $G$ was held fixed)
\begin{equation}\label{smarr-eta3}
    (D-3)M=(D-2)\frac{\kappa A}{8\pi G}-\frac{\Theta \Lambda}{4\pi G}+\sum_i[(D-2)\alpha_i+\beta_i]G^{\gamma_i}\tau_i\Phi_i\, ,
\end{equation}
where $A_i=G^{\gamma_i-\alpha_i+1}\Phi_i$.
\item Endowed with these quantities, one uses the varying-$G$ map of \ref{def:dict}, in order to find the transformation to the holographic first law and Smarr relation given by Eqs.\eqref{1stlaw2} and \eqref{smarr2}.
\end{enumerate}

We shall employ this algorithm to the following examples.


\subsection{Charged, rotating black hole}\label{subsec:qjm}
Consider the charged, rotating black hole, such that the charges considered are the electric charge $\tau_1=Q$ and angular momentum $\tau_2=J$. By referring to the Kerr-Newman metric \cite{Poisson:2009pwt}, we see that the electric charge couples to the gravitational constant as $GQ^2$, which means that from the definition of the quantity $\eta_1$ (referred in the second item of the above algorithm) it would be natural to define $\eta_1=\sqrt{G}Q$. From this prescription, one sees that the power of the gravitational constant coupling is $\alpha_1=1/2$. On the other hand, the electric charge itself has length dimensions $[Q]=l^{(D-4)/2}$, which means that $\beta_1=(D-4)/2$ (this behavior was verified in \cite{Cong:2021fnf} for the Kerr-Newman case and in \cite{Alfaia:2021cnk} for the Reissner-Nordstr\"{o}m one).
\par
For the angular momentum, to find its length dimension $l$, we could rely on the Kerr parameter $a=J/M$ that has obeys $[a]=l$. Since $[M]=l^{-1}$, we should have $[J]=l^0$, thus giving $\beta_2=0$. To verify its $G$-coupling, we recover $G$ in the thermodynamic quantities of the Kerr parameter $a=GJ/(GM)$. Therefore, the quantity $\eta_2=GJ$, which gives $\alpha_2=1$.
\par
The form of the potentials $A_i$ can be seen from the form of the first law of black hole mechanics and Smarr relation for the quantities $\eta_i$ \eqref{1law-eta2}, \eqref{smarr-eta2} and \eqref{smarr-eta3}. This leads to
\begin{align}
d(GM)&=\frac{\kappa}{8\pi}dA+\frac{\Theta}{8\pi} d\Lambda+\sqrt{G}\Phi d\left(\sqrt{G}Q\right)+\Omega d(GJ)\, ,\\
(D-3)M&=(D-2)\frac{\kappa A}{8\pi G}-\frac{\Theta \Lambda}{4\pi G}+(D-3)Q\Phi+(D-2)J\Omega\, .
\end{align}

Notice that since $\sqrt{G}$ is absorbed into the electric charge variation, one must have $A_1=\sqrt{G}\Phi$, where $\Phi$ is the electric potential. And since $G$ is coupled to $J$ in the variation equation, we must have $A_2=\Omega$, the angular velocity. With this quantities at hand, we can use the varying-$G$ map \ref{def:dict} to find the redefinition \eqref{nui}, \eqref{bi}
\begin{align}
 \nu_1&=\tilde{Q}=QG^{-1/2}L\, ,\quad B_1=\tilde{\Phi}=\sqrt{G}\Phi/L\, ,\\
 \nu_2&=J\, ,\quad B_2=\Omega\, .
\end{align}

Therefore, from \ref{theo:1law}, the holographic first law and Smarr relation read
\begin{align}
 dE&=TdS+\tilde{\Phi}d\tilde{Q}+\Omega dJ-pd{\cal V}+\mu dC\, ,\\
 E&=TS+\tilde{\Phi}\tilde{Q}+\Omega J+\mu dC\, ,
\end{align}
which recovers the results of \cite{Cong:2021fnf}.


\subsection{Charged black hole with an alternative convention for the charge coupling}
If one considers different conventions for the bulk charges, i.e., such that they grow differently with $G$, the dictionary would simply imply in a different translation. For instance, considering the convention of \cite{Cong:2021jgb} (see Eqs.(2.3) and (2.5), and notice that $d=D-1$) one has $q\propto\eta_1=GQ$, which implies in $\tau_1=Q$, and $\alpha_1=1$. By referring to the metric function (2.3) of \cite{Cong:2021jgb} and that $[G]=l^{D-2}$, we see that $[\eta_1]=l^{D-3}\Rightarrow [Q]=l^{-1}$, i.e., $\beta_1=-1$. For this reason, since $G$ couples to $Q$ linearly, we identify the shape of the potential $A_1=\Phi$. In fact, we check it by looking at the first law and Smarr relation with variable $G$
\par
\begin{align}
 d(GM)&=\frac{\kappa}{8\pi}dA+\frac{\Theta}{8\pi} d\Lambda+\Phi d\left(GQ\right)\, ,\\
(D-3)M&=(D-2)\frac{\kappa A}{8\pi G}-\frac{\Theta \Lambda}{4\pi G}+(D-3)Q\Phi\, .
\end{align}
\par
From these parameters ($\alpha_1=1$, $\beta_1=-1$), the use of the varying-$G$ map \ref{def:dict} leads to the following redefinition, followed by the holographic first law and Smarr relation:
\begin{align}
 v_1&=\tilde{Q}=QL\, ,\quad B_1=\Phi/L\, ,\\
 dE&=TdS+\tilde{\Phi}d\tilde{Q}-pd{\cal V}+\mu dC\, ,\\
 E&=TS+\tilde{\Phi}\tilde{Q}+\mu dC\, ,
\end{align}
which coincides with the results of \cite{Cong:2021jgb}, given by Eqs.(2.20)-(2.23) (for $R=L$, where $R$ is the radius of the sphere in which the CFT is resides).


\subsection{Kiselev black hole}
Consider the case of the Kiselev black hole \cite{Kiselev:2002dx,Alfaia:2021cnk}, whose static metric function is given by 
\begin{equation}
    f(r)=1-\frac{16\pi GM}{(D-2)\Omega_{D-2}r^{D-3}}-\frac{2\Lambda r^2}{(D-1)(D-2)}-\frac{G\,  b}{r^{(D-1)\omega+D-3}}\, .
\end{equation}

In this case, we have $\eta_1=Gb$ and $\tau_1=b$, such that $[\eta_1]=l^{(D-1)\omega+D-3}$. This means that the dimension of $\tau_1$ is $[\tau_1]=l^{(D-1)\omega-1}$ (since $[G]=l^{D-2}$). Therefore, we have $\alpha_1=1$, $\beta_1=(D-1)\omega-1$. From this, the first law and Smarr relation read
\begin{align}
d(GM)&=\frac{\kappa}{8\pi}dA+\frac{\Theta}{8\pi} d\Lambda+B d\left(Gb\right)\, ,\\
(D-3)M&=(D-2)TS-\frac{\Theta \Lambda}{4\pi G}+[\omega(D-1)+D-3]Bb\, .
\end{align}
Therefore, from $A_1=B$, and we can apply the varying-$G$ map \ref{def:dict} to redefine the thermodynamic quantities and find the holographic first law and Smarr relation as follows
\begin{align}
 \nu_1&=\tilde{b}=bL^{1-(D-1)\omega}\, ,\quad B_1=\tilde{B}=BL^{(D-1)\omega-1}\, ,\\
 dE&=TdS+\tilde{B}d\tilde{b}-pd{\cal V}+\mu dC\, ,\\
 E&=TS+\tilde{B}\tilde{b}+\mu dC\, .
\end{align}

In fact, whenever a quantity couples linearly to $G$, such that $\eta_i=G\tau_i$, i.e., if $\alpha_i=1$, then, the varying-$G$ map simply gives $\nu_i=\tau_i L^{-\beta_i}$ and  $B_i=A_i L^{\beta_i}$, implying that $\nu_i$ becomes dimensionless.


\subsection{AdS Taub-NUT black hole}
This is a $4$-dimensional vacuum axisymmetric solution of Einstein's equations with a negative cosmological constant whose simplest metric is given by
\begin{equation}
    ds^2=-f[dt+(2n\cos \theta+s)d\phi]^2+\frac{dr^2}{f}+(r^2+n^2)(d\theta^2+\sin^2\theta d\phi^2)\, ,
\end{equation}
where $f=(r^2-2Mr-n^2)/(r^2+n^2)-(3n^4-6n^2r^2-r^4)/(L^2(r^2+n^2)$, the parameter $n$ is allowed to vary and is interpreted as a magnetic charge analog to the mass and $s$ is a parameter related to the strength of a singular string presented in this spacetime, called Misner string \cite{BallonBordo:2021pzm,Papadimitriou:2005ii}.

Now let us follow our proposed algorithm by starting with step 1, in which we search for the relevant charges $\tau_i$ that will be redefined. Usually, by placing $G=1$, the  first law of extended black hole thermodynamics would read (see Eq.(4.144) of \cite{BallonBordo:2021pzm})
\begin{equation}
    dM=TdS+V dP +\phi_{+}dN_{+} +\phi_{-}dN_{-}\, ,
\end{equation}
where $N_{\pm}$ are charges related to the variation of $n$ and $s$. In order to allow $G$ to vary and start using the algorithm described in this paper, we write the first law as
\begin{equation}\label{taub-1st1}
 d(GM)=\frac{\kappa}{8\pi }dA+\frac{\Theta}{8\pi} d\Lambda +\sum_i A_id(G^{\alpha_i}\tau_i)\, ,
\end{equation}
where $i=\{1,2\}$, $A=\pi(r_+^2+n^2)$, $\kappa=[1+3(n^2+r_+^2)/L^2]/(2r_+)$, $\Lambda=-3L^{-2}$, $\Theta=-V=-4\pi r_+^3(1+3n^2/r_+^2)/3$, $\tau_1=N_+$ and $\tau_2=N_-$ (and $r_+$ is the {\it locus} of the outer horizon). In this case, the powers $\alpha_i$ and the conjugate quantities $A_i$ will be found in the next steps.
\par
For step 2, we recover $G$ in the metric functions in order to identify the powers $\alpha_i$. Since $f$ is a dimensionless function, the only place where $G$ appears is as a factor of $M$, therefore $n$ (and consequently $s$) is a parameter with dimensions of length that does not carry any factor with powers of $G$. In fact, the charges $N_{\pm}$ read
\begin{equation}\label{npm}
    N_{\pm}=\frac{-2\pi n(n\pm s)^2}{r_+}\left(1+\frac{3(n^2-r_+^2)}{L^2}\right)\, ,
\end{equation}
which means that when we bring back $G$ to the equations, these quantities would not get coupled with it (this is the simplest case), which implies that $\alpha_i=0$.
\par
For step 3, we simply check from Eq.\eqref{npm} that $[N_{\pm}]=l^2$, which means that $\beta_i=2$.
\par
As steps 4 and 5, we see that the first law read \eqref{taub-1st1} with $\alpha_i=0$, which means that from the form of the Smarr relation in standard extended phase space thermodynamics (see, for instance, Eq.(4.139) of \cite{BallonBordo:2021pzm})
\begin{equation}
    M=2\left(\frac{\kappa A}{8\pi G}-\frac{\Theta \Lambda}{8\pi G} +\psi_+N_+ +\psi_-N-\right)\, ,
\end{equation}
we must have $A_{1,2}=G \psi_{\pm}$, implying that $\gamma_i=0$. With the identification of the parameters $(\alpha_i,\beta_i,\gamma_i,D)=(0,2,0,4)$ and of the relevant charges and potentials, the transformation described in Eqs.\eqref{energy1}-\eqref{mu1} gives
\begin{align}
    &\nu_{1,2}=G^{-1}\tau_{1,2}=G^{-1}N_{\pm}=\tilde{N}_{\pm}\, ,\\
    &B_{1,2}=A_{1,2}=G\psi_{\pm}=\tilde{\psi}_{\pm}\, ,\\
    &dE=TdS-pd{\cal V}+\tilde{\psi}_+d\tilde{N}_++\tilde{\psi}_-d\tilde{N}_-+\mu dC\, ,\\
    &E=TS+\tilde{\psi}_+\tilde{N}_++\tilde{\psi}_-\tilde{N}_-+\mu C\, .
\end{align}
From these redefinitions, it is possible to study the thermodynamics of this kind of black hole in this novel context.

\subsection{Casimir energy and Kerr-AdS$_5$ black hole}
Now, we analyze the specific role that Casimir energy plays in this context. We have been analyzing the case in which the energy considered in the first law is given by the one of a ground state, i.e., the vacuum energy of the CFT was effectively considered as null. However, there exists an energy for the ground state, or Casimir energy, for CFTs on a curved background, as argued in the Appendix E of Ref. \cite{Visser:2021eqk}. However, we can still derive a first law of black hole thermodynamics even in this case, but with the CFT energy $E$ identified with the black hole ADM mass up to a constant $E_{\text{ren}}=M+M_{\text{Cas}}$, where $M_{\text{Cas}}$ corresponds to the Casimir energy of the dual CFT (see, for instance, Appendix E of \cite{Visser:2021eqk}). The effect of such energy addition in the standard extended phase space thermodynamics is a renormalization of the Euclidean ``volume'' as
\begin{equation}
    \Theta_{\text{ren}}=-\frac{\Omega_{D-2}}{D-1}\left(r_+^{D-1}-\frac{D-3}{D-2}L^{D-1}M_{\text{Cas}}\right)\, .
\end{equation}

And the effect in the holographic approach described in this paper consists in a renormalization of the chemical potential $\mu$ due to the renormalized Smarr relation as
\begin{equation}
    E_{\text{ren}}=TS+\mu_{\text{ren}}C+\sum_i B_i\nu_i\, ,
\end{equation}
and of the holographic pressure $p$ according to \eqref{v1}
\begin{equation}
    p_{\text{ren}}=\frac{E_{\text{ren}}}{(D-2){\cal V}}\, .
\end{equation}

As an example of this subject, we consider the case of the Kerr-AdS$_5$ black hole, which as has been shown, for instance in \cite{Papadimitriou:2005ii}, to present such renormalized energy, with a first law given by Eqs.(6.55) of \cite{Papadimitriou:2005ii} and Eq.(2.38) of \cite{Kubiznak:2016qmn}
\begin{equation}
    dM_{\text{ren}}=TdS+\Omega_a dJ_a+\Omega_b dJ_b+VdP\, ,
\end{equation}
where $M_{\text{ren}}=M-M_{\text{Cas}}$, $\Omega_{a,b}$ are angular velocities relative to a non-rotating frame at infinity with rotation parameters $a$ and $b$, $J_{a,b}$ are the corresponding angular momenta charges and  $M_{\text{Cas}}$ is a function of the gravitational constant $G$, AdS radius $L$ and of the rotation parameters $a$ and $b$.

Since the angular momenta continue to have length dimensions $0$ (as discussed in the first case analyzed in this section, \ref{subsec:qjm}) the results of this subsection can be translated from subsection \ref{subsec:qjm}, and considering the addition of an extra angular momentum charge and the renormalization of the energy, which in its turn, reflects in a renormalization of the chemical potential. This way, the holographic first law and Smarr relation are
\begin{equation}
    dE_{\text{ren}}=TdS+\Omega_adJ_a+\Omega_bdJ_b-p_{\text{ren}}d{\cal V}+\mu_{\text{ren}}dC\, ,
\end{equation}
where 
\begin{equation}
    E_{\text{ren}}=TS+\Omega_aJ_a+\Omega_bJ_b+\mu_{\text{ren}}C\, .
\end{equation}




\section{Concluding remarks}\label{sec:conc}

We defined a general sequence of steps with which one can translate quantities that are usually considered in standard asymptotically AdS black hole extended phase space thermodynamics, i.e., the one in which the cosmological constant is allowed to vary, to those that are considered in CFTs in the boundary of the AdS space. Such a construction aims to serve as a guide for future analyses carried out in different scenarios on the recently considered proposal, for instance described in \cite{Cong:2021fnf,Visser:2021eqk,Cong:2021jgb}, to make compatible the black hole thermodynamics in AdS space and the thermodynamics defined in its boundary in light of the holographic principle. Our goal is that the algorithm proposed in this paper can be applied for a more general class of asymptotically anti-de Sitter black holes, thus working as a shortcut when one aims to investigate this subject in different scenarios.


\section*{Acknowledgements}

I. P. L. was partially supported by the National Council for Scientific and Technological Development - CNPq grant 306414/2020-1 and by the grant 3197/2021, Para\'iba State Research Foundation (FAPESQ). I. P. L. would like to acknowledge aslo the contribution of the COST Action CA18108. J.P.M.G. is supported by Conselho Nacional de Desenvolvimento Cient\'ifico e Tecnol\'ogico (CNPq) under Grant No. 151701/2020-2. 
H.B.-F. is partially supported by Conselho Nacional de Desenvolvimento Cient\'ifico e Tecnol\'ogico (CNPq) under Grant No. 311079/2019-9. This study was financed in part by the Coordena\c c\~ao de Aperfei\c coamento de Pessoal de N\'ivel Superior (CAPES), finance code 001.

\bibliographystyle{utphys}
\bibliography{hol-dict}

\end{document}